\documentstyle[sprocl]{article}

\input{epsf}
\newcommand{\simle}{\mbox{\,\raisebox{.5mm}{$<$}\hspace{-3.8mm}\raisebox{-1.2mm}{
$\sim$}}\,}

\def\be{\begin{equation}}
\def\ee{\end{equation}}
\def\bea{\begin{eqnarray}}
\def\eea{\end{eqnarray}}

\begin{document}

\title{Intermittency in Plasma Turbulence}

\author{V. Carbone, P. Giuliani, L. Sorriso--Valvo, P. Veltri}
\address{Dipartimento di Fisica, Universit\'a della Calabria, 87036 
Roges di Rende, Italy and \\ 
Istituto Nazionale per la Fisica della Materia, Unit\'a di Cosenza, Italy}

\author{R. Bruno}
\address{Istituto di Fisica dello Spazio Interplanetario/CNR c.p. 27, 00044 
Frascati, Italy}

\author{E. Martines$^1$, V. Antoni$^{1,2}$}
\address{$^1$Consorzio RFX, Corso Stati Uniti, Padova, Italy \\
$^2$Istituto Nazionale per la Fisica della Materia, Unit\'a di Padova, Italy}

\maketitle\abstracts{Intermittency in fluid turbulence can be evidentiated 
through the analysis
of Probability Distribution Functions (PDF) for velocity fluctuations,
which display a strong non--gaussian behavior at small scales. In this
paper we investigate the occurrence of intermittency in plasma turbulence
by studying the departure from the gaussian distribution of PDF for
both velocity and magnetic fluctuations. We use data coming from two
different experiments, namely {\it in situ} satellite observations of the
inner solar wind and turbulent fluctuations in a magnetically confined fusion
plasma. Moreover we investigate also time intermittency observed in a
simplified shell model which mimics 3D MHD equations. We found that the
departure from a gaussian distribution is the main characteristic of all
cases. The scaling behaviour of PDFs are then investigated by using two 
different models built up in the past years, in order to capture the essence 
of intermittency in turbulence.}

\section{Introduction}

The statistics of turbulent fluid flows can be characterised by the Probability 
Distribution Function (PDF) of velocity differences 
$\delta u_r = u(x+r)-u(x)$ over varying scales $r$ (see \cite{frisch} and 
references therein). At large scales the PDF is 
approximately gaussian, as the scale decreases, the wings of the distribution 
become increasingly stretched, so that large deviations from the average value 
are present. This phenomenon is usually ascribed to intermittency. 
The way the PDF departs from the gaussian is very interesting because this 
has to do with models for intermittency in turbulence. If we introduce a 
scaling for fluctuations, namely $\delta u_r \sim r^h$ as the fluid equations 
indicate, a change of scale $r \to l r$ (with $l > 0$) leads to 

\[
\delta u_{l r} = l^h \delta u_r
\]
This is interpreted as an equality in law \cite{frisch}, that is the 
right--hand--side has the same statistical properties as the left--hand--side. 
It can be easily shown that if $h$ is unique, that is in a pure self--similar 
(fractal) situation, 
the standardized variables $\delta u_r/\langle\delta u_r^2\rangle^{1/2}$ 
have the same PDF for all scales $\tau$. 
On the contrary me must invoke the multifractal theory (or some other 
models) to describe intermittency, by conjecturing the presence of an
entire range of values of $h$.

Intermittency has been observed some time ago and deeply investigated in fluid
flows \cite{frisch}, and more recently also in Magnetohydrodynamic (MHD) flows
(see for example \cite{biskamp}). 
In particular, experimental studies carried out in MHD flows
deal mainly by analyses of satellite measurements of solar wind fluctuations
\cite{burlaga91,marscheliu93,carbone95,carbone96,ruzy95,horbury97}, or by
using high resolution 2D numerical simulations \cite{politano,2dmhd} and
Shell Models \cite{carbone94,giuliani98}. All these analysis deal with the
scaling exponents of structure functions, aimed to show that they follow
anomalous scaling laws which can be compared with the usual energy cascade
models for turbulence.

The non gaussian nature of PDF in MHD solar wind turbulence has been 
evidenced by Marsch and Tu \cite{marschetu94}, followed by quantitative 
analysis made by Sorriso--Valvo et al. \cite{noi99}. In fact, in order to 
investigate the properties of intermittency through the 
analysis of non gaussian character of PDF, it would be necessary to quantify 
the departure of PDF from gaussian statistics and to analyse how this 
departure depends on the scale $\tau$. In this work we present two models, 
introduced earlier to describe intermittency in fluid flows. 
Both models describe the evolution of PDF with $\tau$, 
by allowing some free parameters to evolve with the scale $\tau$. 
We then analyse some data of plasma turbulence coming from very
different situations. First of all we show that both models are able to
capture time intermittency observed in a discrete model which describes the
MHD turbulent cascade (Section 3), then we show that also intermittency in
real MHD turbulence, say the turbulence observed in the Solar Wind (Section 4)
and magnetic turbulence as observed in a fusion device (Section 5), is
described by the models. We will describe intermittency in plasma
turbulence through the scaling laws of the characteristic parameters of both
models. Finally (Section 6) we discuss the results we have obtained. 

\section{Models for PDF}

As stated above, in this section we briefly introduce two models which 
describe the scaling evolution of PDF, trying to describe intermittency 
through the scale variation of characteristic parameters. 
We consider the model 
based on a log--normal energy cascade proposed by Castaing et al. 
\cite{castaing,vassilicos}; the second model describes the PDF through a 
sequence of stretched exponentials and can be formally derived through the 
theory of extreme deviations \cite{sornette97}.

\subsection{The energy cascade model}

Because of the idea of self--similarity underlying the energy cascade 
process in turbulence, Castaing and co--workers \cite{castaing} introduced a 
model which tries to characterize the behaviour of the PDFs 
through the scaling law of a parameter describing how the shape of the PDF 
changes in going towards small scales \cite{vassilicos}. In its 
simplest form the model can be 
introduced by saying that the PDF of the increments $\delta \psi$ 
(representing here both velocity and magnetic fluctuations) at a given scale 
$\tau$, is made by a convolution of the typical Gaussian distribution 
$P_G(\delta \psi, \sigma) = 
(\sqrt{2\pi}\sigma)^{-1}\exp{(-\delta\psi^2/2\sigma^2)}$,               
with a function $G_{\tau}(\sigma,\lambda)$ which represents the 
weight of the gaussian distribution characterized by the standard deviation 
$\sigma$:

\begin{equation}
P_{\tau}\left(\delta \psi,\lambda\right) = 
\int G_{\tau} \left(\sigma,\lambda \right) 
P_G\left(\delta \psi, \sigma \right) d \sigma \;,
\label{equ1}
\end{equation}
where $\lambda$ is a parameter representing the width of $G_{\tau}$.  

In the usual approach where the energy cascade 
is introduced through a fragmentation process, $\sigma$ is directly related 
to the local energy transfer rate $\epsilon$. In a self--similar situation, 
where the energy cascade generates only a scaling variation of $\sigma = 
\langle\delta \psi^2\rangle^{1/2}$ 
according to the classical Kolmogorov's picture 
\cite{frisch}, $G_{\tau}(\sigma)$ reduces to a Dirac function 
$G_{\tau}(\sigma) = \delta (\sigma-\sigma_0)$. In this case from eq. 
(\ref{equ1}) a Gaussian distribution $P_{\tau}(\delta \psi) = 
P_G(\delta \psi, \sigma_0)$ is retained. 
On the contrary when the cascade is not strictly self--similar, the width of 
the distribution $G_{\tau}$ is different from zero. In this way the scaling 
behaviour of the parameter $\lambda$                                %
(which takes into account the height of the PDFs wings)            
can be used to characterize intermittency.

\subsection{A model coming from Extreme Deviations Theory}

Recently Frisch and Sornette \cite{sornette97}, have built up a more general 
fragmentation model which is based on the extreme deviations theory (EDT), 
which yield non--gaussian PDF as the number of fragments increases. Let us 
consider a series of random events $x_i$, independent and distributed 
according to a probability density $p(x)$. According to the central limit 
theorem, the sum $y = \sum_{i=1}^n x_i$
is distributed according to a gaussian function, that is fluctuations $y/n$ 
are $O(n^{-1/2})$. On the contrary EDT deals with events where 
fluctuations are $O(1)$, with a finite value $n$ of fragments and large 
values of $y$. The theory allows us to recover the asymptotic shape of the 
PDF $P_n(y)$ starting from $p(x)$. The authors \cite{sornette97} consider the 
general case where $p(x) = \exp[-f(x)]$, and $f(x)$ is differentiable and 
increases quite rapidly as $x \to \infty$. Assuming that the second derivative 
is positive $f^"(x) > 0$ and that $x^2 f^"(x) \to \infty$ (for large $x$), 
the asymptotic form for the PDF of $y$ is given by

\[
P_n(y) \sim \exp[-n f(x/n)] \; .
\]
The probability conditioned on $x_i$ whose sum is $y$, is localized for 
large $y$ in a ``democratic" manner near the value for which $y/n \simeq x_i 
\simeq x_2 \simeq ... \simeq x_n$. If during the fragmentation process the 
fragments have probability $p(x) \sim \exp(-C x^{\gamma})$ with $\gamma > 0$, 
and $y = \log S$, where $S = w_1 w_2...w_n$ (that is $x_i = \log w_i$), 
then

\[
P(S) \sim \exp(-C n S^{\gamma/n}) \; .
\]
From the turbulence point of view, if at the scale $\tau$ the value of each 
$\delta \psi$ is due to the product of fragments derived from structures at 
larger scale, from EDT we must expect a PDF as a stretched exponential

\begin{equation}
P(\delta \psi) \sim \exp \left[ -\beta(\tau) |\delta \psi|^{\mu(\tau)} 
\right]
\label{stretched}
\end{equation}
where $\beta(\tau)$ and $\mu(\tau)$ are the parameter we must fix from 
data.

With these models in mind we will discuss now what kind of results are 
obtained for plasma turbulence.

\section{A shell model for MHD turbulence}

Turbulence is a phenomenon in which chaotic dynamics and power law statistics
coexist. In general, due to the high Reynolds numbers involved in realistic
situations, direct numerical simulations of 3D turbulence is not possible. In
order to avoid this problem and in order to capture the gross features of
turbulence, an alternative approach can be followed by investigating reduced
models of turbulence, such as shell models \cite{bohr}. In these models the
wave vector domain is spaced exponentially through $k_n = k_0 2^n$ ($n = 1,
..., N$) and the nonlinear dynamics is reproduced by using a single complex
variable for each $n$--th shell, that is $u_n(t)$ for the velocity field and
$b_n(t)$ for the magnetic field. The quadratic nonlinear  interactions between
modes are then recovered by imposing that: i) the dynamical equations for each
shell variable must contain the  same quadratic terms of the original
equations; ii) the quadratic  invariants of equations must be correctly
preserved in the inviscid  case; iii) the nonlinear coupling occur between
nearest or next nearest shells, modeling local interactions in the $k$ space.
By using these criteria an MHD  shell model which is the analogous of the GOY
model for hydrodynamics  is easily obtained (see \cite{giuliani98} and
\cite{giuliani2} for a complete  review on shell models):    

\begin{eqnarray} 
{du_n \over dt} &=& i k_n \left [ (u_{n+1} u_{n+2} - b_{n+1} b_{n+2}) 
  - {1 \over 4}(u_{n-1} u_{n+1} - b_{n-1} b_{n+1}) + \right.\\ 
\nonumber 
&-& \left. {1 \over 8}(u_{n-2} u_{n-1} - b_{n-2} b_{n-1}) \right]^* 
- \nu k_n^2 u_n + f_n 
\label{shellu} 
\end{eqnarray} 
 
\begin{eqnarray} 
{db_n \over dt} &=& i k_n \left [ {1 \over 6}(u_{n+1} b_{n+2} - b_{n+1} 
u_{n+2}) 
+{1 \over 6}(u_{n-1} b_{n+1} - b_{n-1} u_{n+1}) + \right.\\ 
\nonumber 
&+& \left. {1 \over 6}(u_{n-2} b_{n-1} - b_{n-2} u_{n-1}) \right]^* 
- \eta k_n^2 b_n + g_n 
\label{shellb} 
\end{eqnarray} 
where $\nu$ and $\eta$ are respectively the kinematic viscosity and the
resistivity, and $f_n$ and $g_n$ are forcing terms. Equations (\ref{shellu})
and (\ref{shellb}), in the inviscid and unforced case, conserve the total
energy   
\[ 
E(t) = {1 \over 2} \sum_{n = 1}^N 
\left( |u_n|^2 + |b_n|^2 \right) 
\] 
the cross--helicity 
 
\[ 
H_c(t) = {1 \over 4} \sum_{n = 1}^N 
Re \left( u_n \cdot b_n^* \right) 
\] 
and the magnetic helicity 
 
\[ 
H(t) = \sum_{n = 1}^N (-1)^n {|b_n|^2 \over k_n} 
\] 
 
Apart for other features, shell models share with original equations scaling
laws of the structure functions $S_p(n) = \langle|x_n|^p\rangle \sim
k_n^{-\zeta_p}$ where $x_n$ is either $u_n$ and $b_n$. Scaling exponents
$\zeta_p$ deviates from the  linear Kolmogorov scaling as a consequence of the
intermittency in the dynamical system. 
 
\begin{figure}
\epsfxsize=10cm    
\centerline{\epsffile{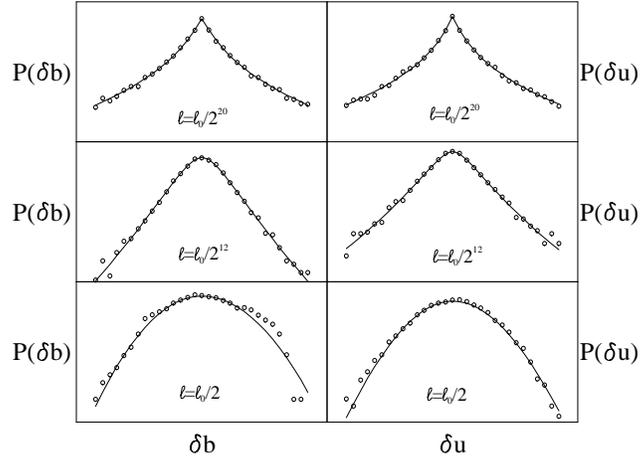}}
\caption{The PDFs of the coefficients $b_n$ and $u_n$ for three different
values of $n$. The fit with the log--normal model by Castaing et al. 
(\ref{equ1}) is superimposed as a full line.}
\label{fig1}
\end{figure}

We solved numerically the system of equations for the shell model, by using  
the following set of parameters: $N = 23$ shells, $\nu = \eta = 10^{-9}$, the  
integration step $dt = 5\times 10^{-5}$. The forcing term acts only on the  
velocity variables (i.e. $g_n=0$) at large scales ($n=1,2$), the magnetic
field being generated by a kind of dynamo effect which is at work in the MHD
shell model \cite{giuliani98}. The forcing, which is assumed to be random,
is calculated according to a Langevin equation $df_n/dt=-f_n\tau_0+\theta$,
where $\tau_0$ is the characteristic time of the largest scales, and $\theta$
is a gaussian white-noise. PDFs for both the velocity and magnetic variables
have been obtained and are shown in figures \ref{fig1} and \ref{fig3} for 
different shells.
As it can be seen PDFs depends on the scale $\ell_n \sim k_n^{-1}$. They are
almost Gaussian at the large scales (small shells $k_n$) and are increasingly
non--Gaussian at small scales (large shells $k_n$). 

In order to make a quantitative analysis of the energy cascade leading to 
the process just described, we have tried to fit the distributions by using, 
for the model (\ref{equ1}), the log--normal ansatz \cite{castaing}

\begin{equation}
G_{\tau} \left(\sigma \right) d\sigma = {1 \over \lambda(\tau)
\sqrt{2\pi} } \exp \left[- {\ln^2 (\sigma / \sigma_0) \over
2 \lambda^2(\tau)} \right] d (\ln \sigma)
\label{lognormal}
\end{equation}
even if also other functions gives rise to results not really different.
The parameter $\sigma_0$ represents the most probable value of $\sigma$,
while $\lambda^2(\tau) = \langle(\Delta \ln \sigma)^2\rangle$ is the width
of the log--normal distribution of $\sigma$.

\begin{figure}
\epsfxsize=8cm    
\centerline{\epsffile{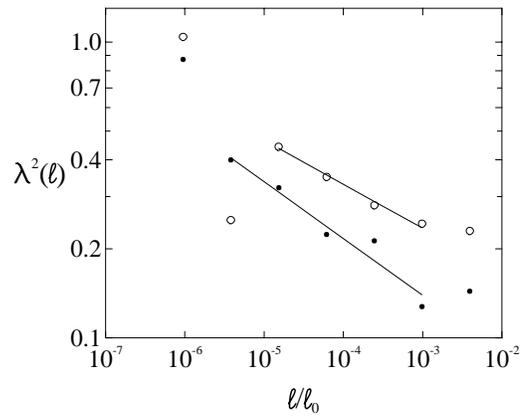}}
\caption{The values of the parameter $\lambda^2$ as a function of the scale
$\ell/\ell_0$. Black circles are for the magnetic field, while the white
circles are for the velocity.}
\label{fig2}
\end{figure}

\begin{figure}
\epsfxsize=10cm    
\centerline{\epsffile{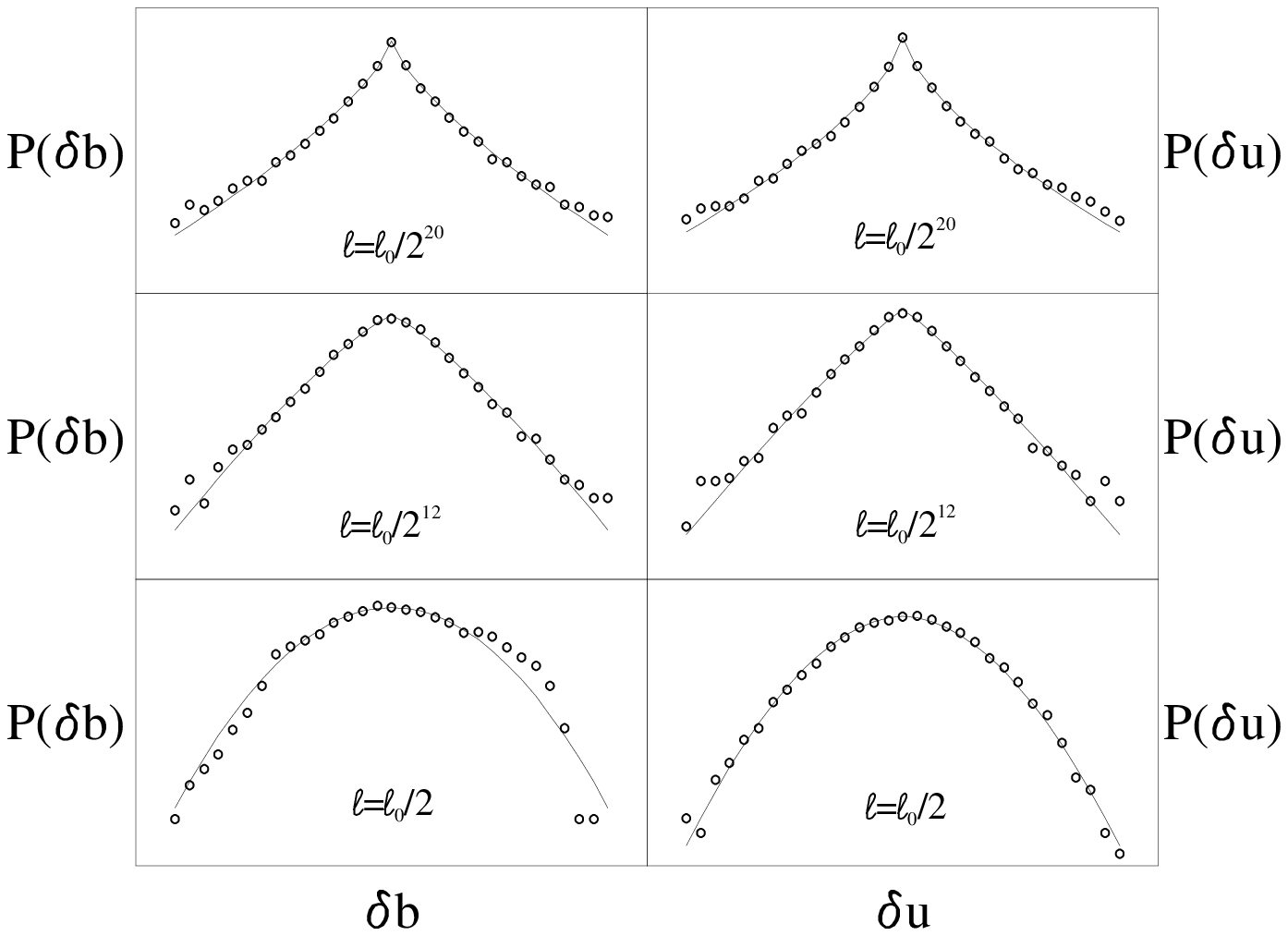}}
\caption{The PDFs of the coefficients $b_n$ and $u_n$ for three different
values of $n$. The fit with the stretched exponentials functions is
superimposed as a full line.}
\label{fig3}
\end{figure}

We have fitted the expression (\ref{equ1}) on the PDFs obtained in the shell
model for both velocity and magnetic intensity, and we have obtained the
corresponding  values of the parameter $\lambda^2$. The values of the
parameters $\sigma_0$, which do not display almost any variation with $\ell_n$
are reported  in table 1. In figure \ref{fig1} we plot,
as full lines, the  curves relative to the fit. As can be seen the scaling
behaviour of PDFs in  all cases is very well described by the function
(\ref{equ1}). From the fit, at each scale $\ell_n$, we get a value for the
parameter $\lambda^2(\ell_n)$, and in figure \ref{fig2} we report the
scaling behaviour of $\lambda^2(\ell_n)$.

\begin{figure}
\epsfxsize=8cm    
\centerline{\epsffile{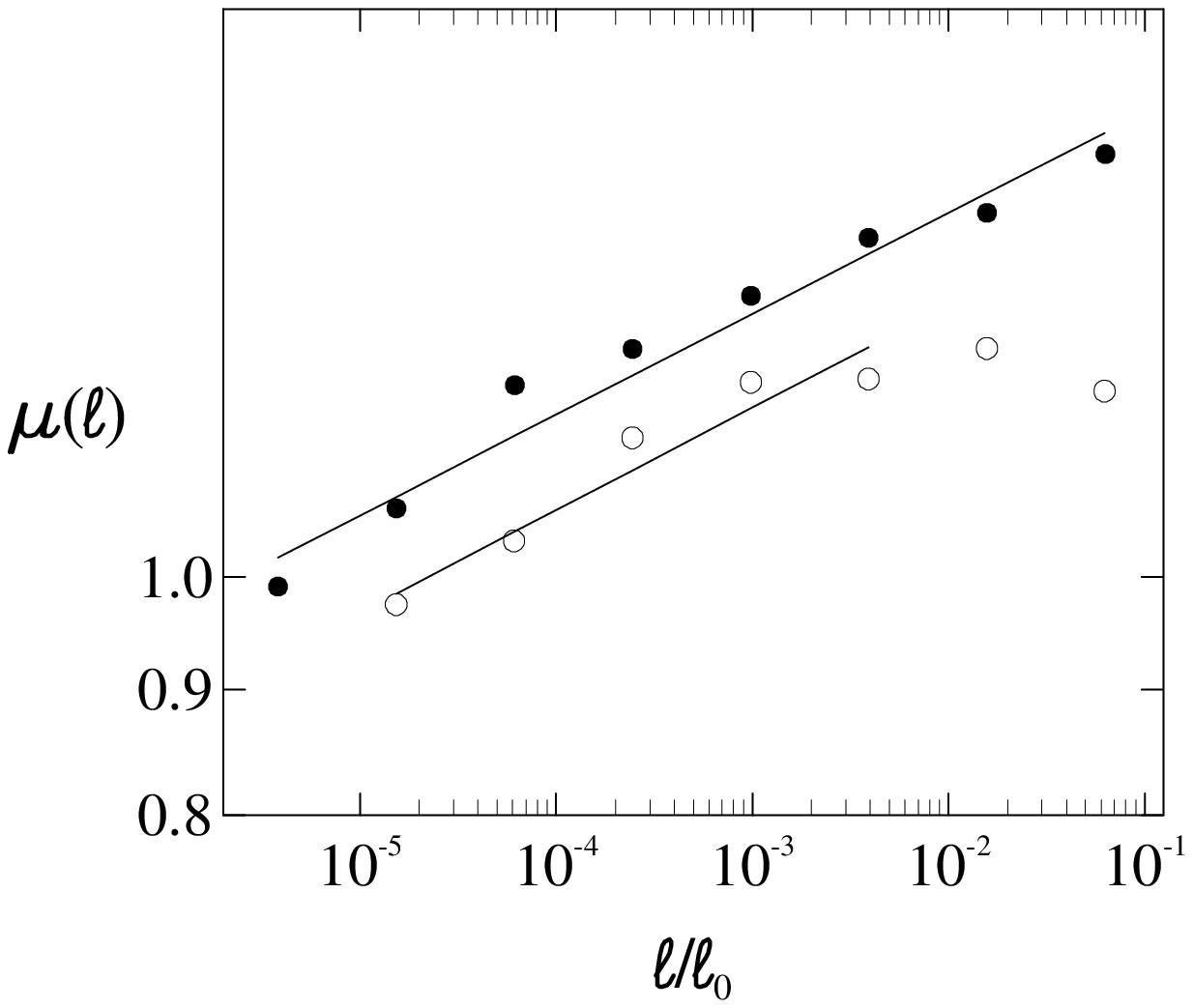}}
\caption{The values of the exponent $\mu$ as a function of the scale
$\ell/\ell_0$. Black circles are for the magnetic field, while the white
circles are for the velocity.}
\label{fig4}
\end{figure}

The scaling evolution of PDFs have been analysed also using the stretched
exponential model $P(\delta x_{\ell}) \sim \exp[-\beta (\delta
x_{\ell})^{\mu}]$; the results of the fit are displayed in figure \ref{fig3}
as a full line. The parameters $\mu$ found from the fit are shown in figure
\ref{fig4} as a function of the scale $\ell_n$, for both $x_{\ell_n}=u_n$ and
$x_{\ell_n}=b_n$. As can be seen, a power law $\mu(\ell_n) \sim \ell_n^a$ is
found in a very wide range for the magnetic field; the range is less wide in
the case of the velocity field, for wich a saturation of $\mu$ is reached. The
minima values for $\mu$ are reported in the table \ref{tabletwo}, together
with the exponents $a$ found for the power laws. 

As we will show in the next two Sections, both models are able to capture
intermittency also for turbulence in non homogeneous situations, namely the
Solar Wind and a fusion plasma. To avoid superpositions we describe for the
Solar Wind only the results obtained with log--normal ansatz, while for the
fusion plasma we describe only the results obtained with the stretched
exponentials, even if in both \ref{tableone} and \ref{tabletwo} the reader can
find the characteristic parameters obtained through the fit of both models for
both samples.

\section{Solar Wind Observations}

The satellite observations of both velocity and magnetic field in the
interplanetary space, offer us an almost unique possibility to gain
information on the turbulent MHD state in a very large scale range, say from 
$1$ AU (Astronomical Unit) down to $10^3$ km. Here we limit to analyse only
plasma measurements of the bulk velocity $V(t)$ and magnetic field intensity
$B(t)$. We based our analysis on plasma measurements as recorded by the
instruments on board Helios 2 during its primary mission in the inner 
heliosphere. The analysis period refers to the first four months of 1976 when 
the spacecraft orbited from $1$ AU, on day 17, to $0.29$ AU on day 108. The 
original data were collected in $81s$ bins and we choose a set of subintervals 
of two days each. The subintervals were selected separately within low speed 
regions and high speed regions selected in a 
standard way according to a threshold velocity \cite{noi99}. For each
subinterval we calculated the velocity and magnetic increments at a given
scale $\tau$ through $\delta V_{\tau} = V(t+\tau)-V(t)$ and $\delta B_{\tau}
= B(t+\tau)-B(t)$, which represent characteristic fluctuations across eddies
at the scale $\tau$. Then we normalize each variable to the standard
deviation within each subinterval $\delta v_{\tau} = \delta V_{\tau}/\langle 
\delta V_{\tau}^2\rangle^{1/2}$ and $\delta b_{\tau} = \delta
B_{\tau}/\langle \delta B_{\tau}^2 \rangle^{1/2}$ (where brackets represent
average within each subinterval at the scale $\tau$).
We calculate the PDFs at 11 different scales logarithmically spaced
$\tau = \Delta t \; 2^n$, where $n=0,1,...,10$ and $\Delta t = 81 s$.
We collect the number of events within each bins by using $31$ bins equally 
spaced in the range within $3$ times the standard deviation of the total 
sample. Before we mixed the different subperiods belonging to a given
class (high or low speed streams), we tested for the fact that the gross
features of PDFs shape does not change in different subintervals. Then our 
results for high and low speed streams are representative of what happens at 
the PDFs.

\begin{figure}
\epsfxsize=8cm    
\centerline{\epsffile{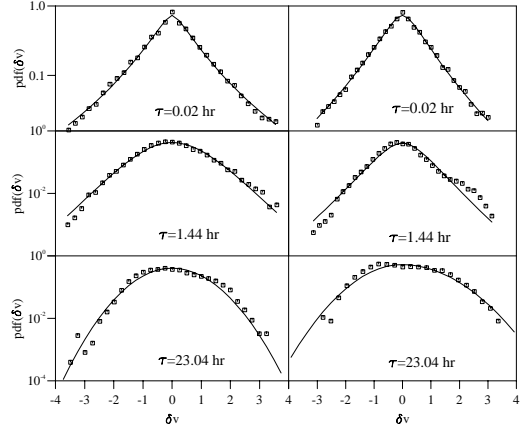}}
\caption{The scaling behaviour of the PDF for $\delta v_{\tau}$ as calculated 
from the experimental data (white symbols) in the fast (left) and slow (right) 
streams. The full lines 
represent the fit obtained through the model by Castaing et al. (\ref{equ1}), 
as described in the text.}
\label{fig5}
\end{figure}

\begin{figure}
\epsfxsize=8cm    
\centerline{\epsffile{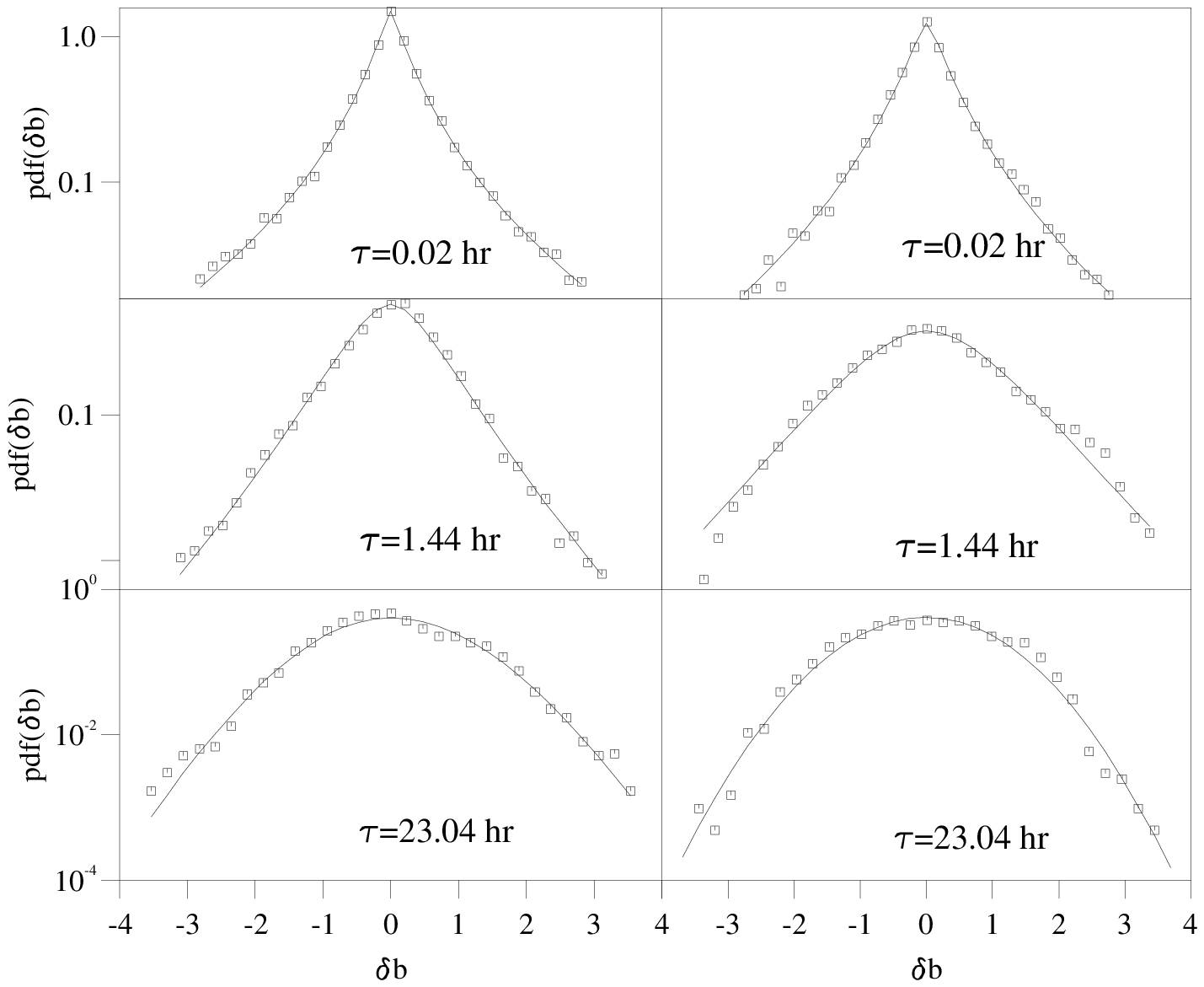}}
\caption{The scaling behaviour of the PDF for $\delta b_{\tau}$ as calculated 
from the experimental data (white symbols) in the fast streams. Again, the 
full lines represent the fit obtained through the model by Castaing et al. 
(\ref{equ1}), as described in the text.}
\label{fig6}
\end{figure}

The results are shown in figures \ref{fig5} and \ref{fig6}, where we
report the PDFs of both velocity and magnetic intensity for both the
high speed streams and the slow speed streams. At large scales the PDFs are
almost Gaussian, and the wings of the distributions grow up as the scale
becomes smaller. This is true in all cases, say for both types of wind. 
Stronger events at small scales have a probability of occurrence greater than
that they would have if they were distributed according to a gaussian 
function. This behaviour is common for intermittency as currently observed in 
fluid flows \cite{frisch} and in the solar wind turbulence 
\cite{marschetu,noi99}.

\begin{figure}
\vspace{-5mm}
\epsfxsize=8cm    
\centerline{\epsffile{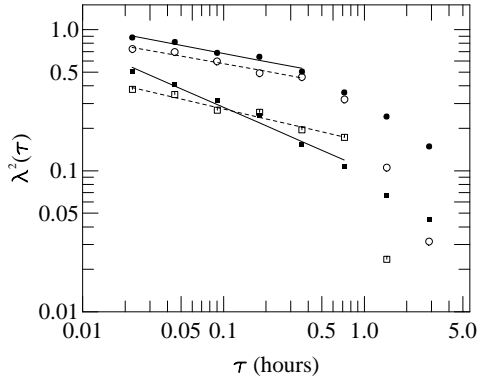}}
\caption{We show the scaling behavior of $\lambda^2(\tau)$ vs. $\tau$ for both 
fast (black symbols) and slow (open symbols) streams. Circles refer to 
the magnetic field intensity, squares refer to the bulk velocity.}
\label{fig7}
\end{figure}

A quantitative analysis of the energy cascade has been made by using the
log--normal ansatz \cite{castaing} (\ref{lognormal}). We have fitted the
expression (\ref{equ1}) on the experimental PDFs for both velocity and
magnetic intensity, and we have obtained the corresponding  values of the
parameter $\lambda^2$. The values of the parameters $\sigma_0$, which again do 
not display almost any variation with $\tau$ are reported in table 1. 
In figures \ref{fig5} and \ref{fig6} we plot, as full lines, 
the curves relative to the fit. 
As can be seen the scaling behaviour of PDFs in all cases is very well
described by the function (\ref{equ1}). From the fit, at each scale $\tau$, we
get a value for the parameter $\lambda^2(\tau)$, and in figure \ref{fig7} we
report the scaling behaviour of $\lambda^2(\tau)$ for both high and low speed
streams. Starting from a very low value at the scales of about $1$ day,
$\lambda^2$ increases abruptly to $\lambda^2 \simeq 10^{-1}$ at about $2$
hours, and finally a scaling law starts to become evident up to $\Delta t =
81$ sec. In this last range, which corresponds roughly to what is usually
called the ``Alfv\'enic range'', we fitted the parameter with a power law
$\lambda^2(\tau) \sim \tau^{-\alpha}$. The values of $\alpha$ obtained in the
fitting procedure and the corresponding range of scales, are reported in the
table \ref{tableone}. In the same table, we report the values of the maximum
value of $\lambda^2(\tau)$ in the range of fit, namely $\lambda^2_{max}$,
which is an indication for the strenght of the intemittency
\cite{2dmhd,noirfx}.

Looking at figure \ref{fig7} and at $\lambda^2_{max}$, it can be seen that
both in fast and in slow streams magnetic field intensity is more intermittent
than bulk velocity (values of $\lambda^2$ are at least two times
larger for magnetic field intensity than for velocity).
The same indications comes from 2D MHD direct simulations
\cite{politano,2dmhd}, and in different analysis of solar wind intermittency
\cite{veltriemangeney}. The values of $\lambda^2(\tau)$ are more or less the
same for magnetic field intensity both in fast and in slow wind.
This is due to the fact that magnetic field intensity fluctuations
are related to compressive fluctuations, which should have the same nature in
both types of wind. The bulk velocity fluctuations on the contrary are more
intermittent at small scales ($81$ sec) in the fast wind and at large scale
($\simeq 1$ hour) in the slow wind. From table \ref{tableone} it appears that
the value of $\alpha$ is not universal, a result which has also been found
in fluid flows \cite{castaing}. The fact that the value of $\alpha$ for the
velocity fluctuations in slow wind is the same as the value of $\alpha$ for
magnetic field intensity suggests that intermittent structures in slow wind
are perhaps mainly associated with compressive fluctuations. On the contrary
the different value of $\alpha$ found in fast wind evidences a different 
nature of velocity fluctuations in fast wind, perhaps related to the fact that
such fluctuations are mainly incompressible. This result is in agreement with
what has been found by Veltri and Mangeney \cite{veltriemangeney}. 
These authors found that in fast wind the more intermittent structures are 
tangential discontinuities with almost no variation in magnetic field 
intensity, while in slow wind
the most intermittent structures are shock waves, which display the same
behaviour in bulk velocity and magnetic field intensity.

\section{The Reverse Field Pinch: RFX experiment}

In this section we report evidences for the presence of intermittency in 
another type of magnetized fluid, namely a plasma of interest for controlled 
thermonuclear fusion research, confined in reversed field pinch (RFP) 
configuration.

The RFP is a toroidal configuration of magnetic fields which is proposed as an
alternative to the tokamak for confining fusion grade plasmas \cite{bodin90}.
The configuration is characterized by toroidal and
poloidal magnetic field components of comparable magnitude (in a tokamak the
field is mainly toroidal). The configuration is a near--minimum energy state 
to which a plasma relaxes under proper constraints \cite{taylor74}.
The toroidal field changes sign in the outer part of the plasma, a feature
which gives the name to the configuration. Such field reversal, which
improves the MHD stability of the configuration, is spontaneously generated
by the plasma, and is maintained against resistive diffusion by the dynamo
process \cite{biskamp}.
This is achieved through the action and nonlinear coupling of
several resistive magnetohydrodynamic modes, which give rise to a high
level of magnetic turbulence (of the order of 1\% of the average field in
present day experiments, i.e. two orders of magnitude larger than in
tokamaks). This high fluctuation level makes the RFP very suited
for the study of MHD turbulence properties, mainly for their magnetic part.
The magnetic turbulence has been demonstrated to be the main cause of
energy and particle transport in the RFP core, whereas at the edge
its contribution is still under investigation. In this region the
electrostatic turbulence has been proved to give an important
contribution to the particle transport \cite{antoni98}. 

The measurements have been obtained in the RFX experiment, which is the 
largest RFP presently in operation (major radius $2$ m, minor radius 
$0.457$ m) \cite{fellin95}. RFX is designed to reach a plasma
current of $2$ MA, and currents up to $1$ MA have been obtained up to now. The
measurements were performed in low currents discharges ($300$ kA) using a
magnetic probe inserted in the edge plasma. The probe consists of a coil
housed in a boron nitride measuring head. The coil measures the time
derivative $\partial_t B$ of the radial component $B(t)$ of the magnetic
field. The radial direction in this case goes from the core plasma to the
edge. The sampling frequency of the measurements is $2$ MHz. Measurements 
have been collected at different values of the probe
insertion $X=r/a$ into the plasma, $X=0$ being the position of the inner
convolution of the RFX graphite first wall, $r$ being the probe radial position 
and $a$ being the radius of the ring. Different measurements are also
identified through a shot number. In RFX two different components of
the magnetic fluctuations can be identified: a localised and stationary
magnetic perturbation, originated by the tearing modes responsible for the
dynamo which tend to be phase--locked and locked to the wall \cite{antoni95}, 
and an high frequency broadband activity, which is investigated here.
All measurements presented were made away from the stationary perturbation.

\begin{figure}
\epsfxsize=8cm    
\centerline{\epsffile{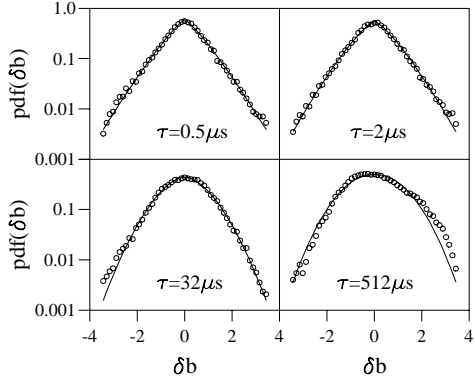}}
\caption{The scaling behaviour of the PDF of the magnetic fluctuation 
$\delta b_{\tau}$ measured in the RFX experiment. The full lines 
represent the fit obtained through the stretched exponential model, 
as described in the text.}
\label{fig8}
\end{figure}

We calculated the characteristic magnetic fluctuations at the scale $\tau$, 
namely $\delta B(\tau) = B(t+\tau)-B(t)$. For each position within the device,
we can study the statistical behaviour of fluctuations at different scales
$\tau$. In figure \ref{fig8} we report the PDFs of 
$\delta b_{\tau} = \delta B_{\tau}/\langle\delta B_{\tau}^2\rangle^{1/2}$ 
at different scales for a given value of $X$. 
As it can be seen the PDFs do not collapse
to a single curve, but follow the usual characteristic scaling behaviour 
evidenced also in solar wind measurements. This behaviour is visible for all 
the values of the insertion $X$. A fit to the PDFs we obtained has been 
made by using a stretched exponential, namely $P(\delta b_{\tau}) \sim 
\exp[-\beta (\delta b_{\tau})^{\mu}]$, and their scaling behaviour is then 
investigated by looking at the scaling
laws of both the parameters $\beta(\tau,X)$ and $\mu(\tau,X)$. 
The parameter $\mu$ follows a power law in a certain range of scales, 
that is: $\mu(\tau,X) \sim \tau^{a(X)}$,
even near the external wall (figure 5). 

\begin{figure}
\epsfxsize=8cm    
\centerline{\epsffile{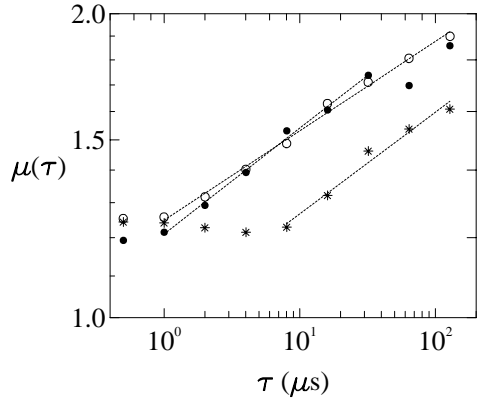}}
\caption{We show the scaling of $c(\tau)$ vs. $\tau$ for the RFX 
 magnetic field at different insertion values: $X=0.97$ (black circles),
 $X=0.95$ (white circles) and $X=0.91$ (stars).}
\label{fig9}
\end{figure}

The scaling exponent $a(X)$ (see table \ref{tabletwo}) remains near to the 
value $0.1$ in the different shots, and the range of scales where a scaling 
for the parameter $\mu(\tau,X)$ is visible is larger for measurements near the 
first wall with respect to measurements inside the device.  
The ``strenght'' of the intermittency effects can be compared for the 
different insertion values $X$, by comparing the minimum values found for the 
exponent $\mu(\tau)$ in each shot. As can be seen from the table 
(\ref{tabletwo}), such values increase going toward the core, showing that the
intermittency is stronger near the wall \cite{noirfx}.

\begin{table}
\caption[Table]{We report the values of the parameters $\sigma_0$,
$\lambda^2_{max}$ and $\alpha$ obtained in the fitting procedure for
$\lambda^2(\tau)$ of the log-normal model in the shell model data, 
solar wind and RFX (shot 8414). 
We also report the range of scales where the fit has been done. 
For the shell model, the range of fit is in $\ell/\ell_0$ units.}
\begin{center}
\begin{tabular}{ccccc}
           & $\sigma_0$ &  $\lambda^2_{max}$ &  $\alpha$  &  Scales (SW: hours) \\
&&&\\[-8pt]
\hline&&&\\[-8pt]
$u_{SM}$        & $0.83 \pm 0.04$ & $0.44 \pm 0.10$ & $0.15 \pm 0.07$ & $1.5\times 10^{-5} \div 10^{-3}$\\
$b_{SM}$        & $0.79 \pm 0.04$ & $0.40 \pm 0.10$ & $0.19 \pm 0.06$ & $3.8\times 10^{-6} \div 10^{-3}$\\
$B_{SW} (fast)$ & $0.85 \pm 0.05$ & $0.88 \pm 0.04$ & $0.19 \pm 0.02$ & $0.02 \div 0.72$ \\
$B_{SW} (slow)$ & $0.90 \pm 0.05$ & $0.73 \pm 0.04$ & $0.18 \pm 0.03$ & $0.02 \div 0.72$ \\
$V_{SW} (fast)$ & $0.90 \pm 0.05$ & $0.51 \pm 0.04$ & $0.44 \pm 0.05$ & $0.02 \div 1.44$ \\
$V_{SW} (slow)$ & $0.95 \pm 0.05$ & $0.37 \pm 0.03$ & $0.20 \pm 0.04$ & $0.02 \div 1.44$ \\
$ RFX_{8414}  $ & $1.01 \pm 0.02$ & $0.21 \pm 0.01$ & $0.42 \pm 0.03$ & $1 \mu s \div 36  \mu s$ \\
[2pt]\end{tabular}
\label{tableone}
\end{center}
\end{table}

\begin{table}
\caption[Table]{The values of $\mu_{min}$ and the scaling exponents $a$,
obtained for the scaling of the exponents $\mu$ of the stretched exponential
fit of the data, from shell model, the RFX and the solar wind. 
The scale ranges in which a power law has been found are also shown. 
In the RFX, different shots atdifferent insertions $X=r/a$ are displayed. 
Two inner shots ($8425$ and $8426$) do not display clear scaling laws. 
However, we report the minima of $\mu$.
For the shell model, the range of fit is in $\ell/\ell_0$ units.}
\begin{center}
\begin{tabular}{ccccc}
                 &  $X$   &   $\mu_{min}$   &      $a$        &  Scales (RFX: $\mu s$)   \\
&&&\\[-8pt]
\hline&&&\\[-8pt]
$u_{SM}$         &  ---   & $0.97 \pm 0.03$ & $0.04 \pm 0.01$ & $1.5\times 10^{-5} \div 3.9\times 10^{-3}$\\
$b_{SM}$         &  ---   & $0.99 \pm 0.03$ & $0.04 \pm 0.01$ & $3.8\times 10^{-6} \div 6.3\times 10^{-2}$\\
$RFX_{8414}$     & $0.97$ & $1.19 \pm 0.02$ & $0.10 \pm 0.01$ & $1 \div 36  $\\
$RFX_{8417}$     & $0.96$ & $1.25 \pm 0.01$ & $0.09 \pm 0.01$ & $1 \div 128 $\\
$RFX_{8420}$     & $0.91$ & $1.23 \pm 0.01$ & $0.10 \pm 0.01$ & $8 \div 128 $\\
$RFX_{8425}$     & $0.90$ & $1.50 \pm 0.02$ &       ---       & ---          \\
$RFX_{8426}$     & $0.86$ & $1.55 \pm 0.03$ &       ---       & ---          \\
$B_{SW} (fast)$  &  ---   & $0.74 \pm 0.01$ & $0.16 \pm 0.01$ &$0.02h \div 24h$\\
$B_{SW} (slow)$  &  ---   & $0.81 \pm 0.02$ & $0.16 \pm 0.02$ &$0.02h \div 24h$\\
$V_{SW} (fast)$  &  ---   & $0.92 \pm 0.02$ & $0.12 \pm 0.02$ &$0.02h \div 24h$\\
$V_{SW} (slow)$  &  ---   & $1.04 \pm 0.03$ & $0.11 \pm 0.01$ &$0.02h \div 24h$\\
[2pt]\end{tabular}
\label{tabletwo}
\end{center}
\end{table}

\section{Discussion and conclusions}

The departure from gaussianity of the PDFs of fields increments is well known
in fluid turbulence, and has been pointed out in MHD plasmas. The data
presented in this paper clearly confirm this peculiar behavior, showing  that
intermittency strongly affects the scaling behavior of MHD turbulence. The
analysis of data with the two models presented here has a duplex interest: i)
it is an experimental test for the theoretical models; ii) it gives a 
quantitative characterization of intermittency in different physical systems,
leading to the the possibility of a comparison between them. 

For the first point, we showed and discussed in detail here the results
obtained when we used the log--normal model to describe intermittency in the
solar wind turbulence, and the stretched exponential model to describe
intermittency in RFX plasma. However it can be shown that both models are
able to capture the scaling behavior of PDFs in both the solar wind and RFX
experiments, just as in the case of the MHD shell model. Very good fits are
obtained, and very close values of the $\chi^2$ are found in both cases
\cite{tesi}. In table \ref{tableone} we report the results relative to the fit
of PDFs obtained in RFX plasma (namely the shot number 8414) with the model by
Castaing et al. \cite{castaing}. Other samples give similar results. In table
\ref{tabletwo} we report the results obtained when we fit the PDFs calculated
with the dataset of the solar wind turbulence, with the stretched exponential
model. 

Concerning the second point, we first observe that the two models give
consistent results. The similar role played by $\lambda^2_{max}$ in the
log--normal model and $\mu_{min}$ in the stretched exponential model, as well
as the slopes of the power-low scaling of $\lambda^2$ and $\mu$, is evident.
We can thus look at such parameters to compare the intermittency of the
magnetic field in both experimental datasets we analysed. As a first comparison
we can observe, looking directely at the PDFs shape of magnetic fluctuations,
that intermittency in solar wind is stronger than in the RFX experiment plasma.
The values of both parameters $\lambda^2_{max}$ and $\mu_{min}$ (see
tables \ref{tableone} and \ref{tabletwo}) confirm such result. Clear
scaling laws are found for the parameters of both models, so that it seems
difficult to prefer one model with respect to the other on the basis of our
analysis. 

Even if the experimental results we presented do not give indications wheter
one model is better than the other in fitting experimental results, the
stretched exponential is a more flexible function with respect to
(\ref{equ1}). This is only a "technical" consideration and does not involve a
theoretical preference. From our experience the code developed for the fit with
the stretched exponential is more performant then the code developed by using
the log--normal model. It is however worthwhile to compare
intermittency found in real datasets with time intermittency found in the
simplified models for MHD turbulence. Looking at the PDFs in figures 
\ref{fig1}, \ref{fig3}, \ref{fig5} and \ref{fig6}, we can
realise that intermittency in the shell model is similar to that obtained in
the solar wind, that is turbulence in the shell model is more intermittent
than turbulence in RFX plasma. Furthermore looking at the values of
$\mu_{min}$, it can be seen that magnetic turbulence in the solar wind is more
intermittent than in the shell model, while for the velocity the intermittency
is comparable in both cases. 

To conclude we would like to remark that numerical
simulations of MHD equations in 2D configurations \cite{2dmhd} display strong
intermittency which has been successfully described by the model introduced by
Castaing et al. \cite{castaing}.

\section*{Acknowledgements} 
We are grateful to H. Rosenbauer and R. Schwenn for
making  the Helios plasma data available to us.

\end{document}